\RequirePackage{graphicx}
\documentclass[aps,prb,floatfix,twocolumn, longbibliography]{revtex4-2}
\usepackage{amssymb}
\usepackage{enumitem}
\usepackage{diagbox}
\usepackage{mathtools}
\usepackage{xcolor}
\usepackage{float}
\usepackage{array}
\usepackage{booktabs}
\usepackage{listings}
\usepackage{lmodern}
\usepackage{mathpazo}
\usepackage{microtype}
\usepackage{graphicx}
\usepackage{natbib}
\usepackage{dcolumn}
\usepackage{bm}
\usepackage{mathtools}
\usepackage{braket}
\usepackage{algorithm}
\usepackage{algpseudocode}
\usepackage[utf8x]{inputenc}
\usepackage{amsmath}
\usepackage{multirow}
\usepackage{amsfonts}
\usepackage{amssymb}
\usepackage{rotating} 

\usepackage{amsthm}
\usepackage{tikz}

\usetikzlibrary{positioning}
\usetikzlibrary{patterns}
\usetikzlibrary{backgrounds}

\theoremstyle{plain}
\theoremstyle{definition}

\newcommand{\Gam}[4]{\hat{a}^\dagger_{#1} \hat{a}^\dagger_{#2} \hat{a}^{}_{#3} \hat{a}^{}_{#4}}

\begin{document}

\title{Accelerated Convergence of Contracted Quantum Eigensolvers through a Quasi-Second-Order, Locally Parameterized Optimization}



\author{Scott E. Smart and David A. Mazziotti}
\email[]{damazz@uchicago.edu}
\affiliation{Department of Chemistry and The James Franck Institute, The University of Chicago, Chicago, IL 60637}%
\date{Submitted May 2, 2022}
%

\begin{abstract}
A contracted quantum eigensolver (CQE) finds a solution to the many-electron Schr{\"o}dinger equation by solving its integration (or contraction) to the 2-electron space---a contracted Schr{\"o}dinger equation (CSE)---on a quantum computer.  When applied to the anti-Hermitian part of the CSE (ACSE), the CQE iterations optimize the wave function with respect to a general product ansatz of two-body exponential unitary transformations that can exactly solve the Schr{\"o}dinger equation. In this work, we accelerate the convergence of the CQE and its wavefunction ansatz via tools from classical optimization theory. By treating the CQE algorithm as an optimization in a local parameter space, we can apply quasi-second-order optimization techniques, such as quasi-Newton approaches or non-linear conjugate gradient approaches. Practically these algorithms result in superlinear convergence of the wavefunction to a solution of the ACSE. Convergence acceleration is important because it can both minimize the accumulation of noise on near-term intermediate-scale quantum (NISQ) computers and achieve highly accurate solutions on future fault-tolerant quantum devices. We demonstrate the algorithm, as well as some heuristic implementations relevant for cost-reduction considerations, comparisons with other common methods such as variational quantum eigensolvers, and a fermionic-encoding-free form of the CQE.
\end{abstract}
\maketitle

\section{Introduction}

The contracted Schr{\"o}dinger equation (CSE),\cite{Mazziotti1998, Alcoba2001, Nakatsuji1976} describes the projection of the molecular Schr{\"o}dinger equation for a $N$-electron system onto a two-electron space, which generates the stationary-state condition of the two-electron reduced density matrix (2-RDM) instead of the wavefunction. Satisfaction of the anti-Hermitian part of the contacted Schr{\"o}digner equation (ACSE) by a quantum state is equivalent to its invariance with respect to all infinitesimal two-body unitary transformations,\cite{Mazziotti2006, Mazziotti2007, Mukherjee2001}  Solution of the ACSE for the 2-RDM was initially challenging because the equation depends on both the 2-RDM and three-electron RDM (3-RDM), making it indeterminate without additional information.  However, the ACSE has been practically solved by reconstructing the 3-RDM by its cumulant expansion as an functional of the 2-RDM\cite{Mazziotti1998_cumulant, Mazziotti2006, Mazziotti2007, Sand2015} and applied to computing strongly correlated ground and excited states in both chemical reactions and conical intersections.\cite{Gidofalvi2009,FoleyIV2011,Valdemoro2009,Alcoba2009}

Recently, the ACSE has been solved on quantum devices with applications to hydrogen chains as well as the benzyne isomers.\cite{Smart2021_prl, Smart2021_benzyne, Boyn2021}  On a quantum computer the ACSE algorithm, known as a contracted quantum eigensolver (CQE), iteratively minimizes the residual of the ACSE in contrast to the variational quantum eigensolvers (VQE) that minimize the energy with respect to parameters according to the Rayleigh-Ritz variational principle.  Instead of propagating only the 2-RDM as in the classical algorithm, we propagate a wave function through state preparation on a quantum computer. Thus, we avoid reconstructed RDMs, as the 2-RDM can be directly measured from the quantum state while the ACSE residual can be directly measured from an auxiliary quantum state.  The resulting ACSE algorithm is a potentially exact RDM approach that scales polynomially in the size of the molecular system.

In this paper we accelerate the convergence of the CQE for the ACSE by developing quasi-second-order algorithms with superlinear convergence.  Convergence acceleration is important for avoiding the accumulation of noise on near-term intermediate-scale quantum (NISQ) computers as well as achieving highly accurate solutions on future fault-tolerant quantum devices.  We draw upon research on optimization algorithms on manifolds,\cite{Taylor1994,Absil2007,Huang2015} which have applications across science and engineering from vision to robotics as well as related algorithms in electronic structure for orbital optimization.\cite{Arute2020,Shepard2015}  We specifically develop and implement a quasi-Newton scheme with the Broyden-Fletcher-Goldfarb-Shanno update and non-linear conjugate gradient algorithms.  The quasi-second-order algorithms avoid storage of the Hessian matrix while providing superlinear convergence. We also demonstrate convergence properties as well as some approximate implementations of the search direction and finally compare the resulting CQE algorithms with classes of common quantum algorithms, including the variational quantum eigensolver.
\section{Theory}

We discuss the solution of the ACSE via CQE in section~\ref{sec:acse}, the local parameterization of the wavefunction in section~\ref{sec:para}, the quasi-second-order accelerations in section~\ref{sec:qso}, and resource optimization in section~\ref{sec:sro}.

\subsection{Solution of the ACSE}

\label{sec:acse}

Given a molecular system with Hamiltonian $\hat{H}$, we can write a contraction of the Schr{\"o}dinger equation onto the two-particle space, known as the contracted Schr{\"o}dinger equation (CSE):
\begin{equation}\label{24CSE}
    \langle \Psi | \hat{H} \Gam{i}{k}{l}{j}| \Psi\rangle = E~ {}^2 D^{ik}_{jl}
\end{equation}
The CSE can be split into a Hermitian and an anti-Hermitian part, the latter of which is called the anti-Hermitian CSE, or ACSE:
\begin{equation}\label{13ACSE}
    \langle \Psi | [\hat{H},\Gam{i}{k}{l}{j} ] | \Psi \rangle = {}^2 A^{ik}_{jl}.
\end{equation}
Here, ${}^2 A^{ik}_{jl}$ is the residual of the ACSE, which is necessarily zero when $|\Psi\rangle$ is an eigenstate of the wave function. We can also obtain the ACSE in Eq~\eqref{13ACSE} by considering unitary transformations $\exp(\epsilon {\hat P})$ generated by a parameter $\epsilon$ and a two-body anti-Hermitian operator $\hat{P}$
\begin{equation}
    \hat{P} = \sum_{ij;kl} {}^{2} P^{ij}_{kl} \Gam{i}{k}{j}{l} .
\end{equation}
The derivative of the energy with respect to the elements of the operator $^{2} P$ yields
\begin{equation}
    \frac{\partial E(\hat{P})}{\partial (^{2} P^{ik}_{jl})} + O(\epsilon^2) = \epsilon \langle\Psi |  [\hat{H},\Gam{i}{k}{l}{j}] | \Psi \rangle  = \epsilon \, {}^2 A^{ik}_{jl},
\end{equation}
which we observe is equal to the residual of the ACSE.  In the solution of the ACSE,\cite{Mazziotti2007} the energy and 2-RDM can be expressed as a system of differential equations in terms of a discretized, time-like parameter $\lambda$ that controls the transformation of the implicit wave function to minimize the energy.  As $\lambda$ increases, we approach a solution of the ACSE. On a quantum computer,\cite{Smart2021_prl} we have a potentially exponential advantage in terms of simulating the exact 2-RDM (or 3-RDM) and ${}^2 A$ matrices.

\subsection{Local Parameterization of the Contracted Quantum Eigensolver}

\label{sec:para}

We can describe the generic problem in the variational quantum eigensolver\cite{McClean2016, Peruzzo2014, Kandala2017, Endo2020} for finding the ground-state wave function of a quantum system as:
\begin{equation}
\min_{\vec{\theta} \in \mathbb{R}^\nu}  \langle \Psi_0| U^\dagger (\vec{\theta} )\hat{H}  U (\vec{\theta})| \Psi_0   \rangle.
\end{equation}
where $\vec{\theta}$ represents a vector of $\nu$ real parameters and we assume the wave function is properly normalized. Upon convergence the following equation is satisfied for all $k$,
\begin{equation}
~\frac{\partial}{\partial \theta_k } \langle \Psi_0| U^\dagger (\vec{\theta} )\hat{H}  U (\vec{\theta})| \Psi_0   \rangle = 0,
\end{equation}
indicating that the gradient with respect to all parameters vanishes. However, there can be significant problems associated with describing the appropriate parameterization of $\Psi$.
The exact solution of the wavefunction scales exponentially, which might imply that an exponential number of parameters is necessary in a variational scheme. Using an operator like unitary coupled cluster provides an exponential ansatz, but because the mapping from the Euclidean space to the unitary space is nonlinear, there can be singularities or unphysical minima in the optimization surface.\cite{Stuelpnagel1964,Evangelista2019} Additionally, it has been shown that high-dimensional parameterizations in a random variational ansatz generate barren plateaus where the variance in the energy gradients vanishes as the system size increases.\cite{McClean2018, Wang2020}
Because an exponential scaling parameterization is not feasible for larger systems, and limited excitation ansatz such as UCC singles and doubles are not sufficiently accurate, a slew of iterative schemes based on the VQE and UCC schemes which deviate from the traditional CC formalism have been proposed, providing scalable approaches that generally repeat or extend upon certain ansatz fragments.\cite{Lee2019, Chen2021, Ryabinkin2018}  One such approach is the adaptive derivative assembled pseudo-trotterized VQE method, or ADAPT-VQE, which takes elements of the ACSE (or generalized UCCSD) to generate an increasingly more complex variational problem.\cite{Grimsley2018}

In the CQE approach, we instead forgo the global parametrization of the state and use an atlas of local parametrizations,\cite{Taylor1994} describing the trajectory of the state, with the parameter space being dependent on the contracted eigenvalue equation (which here is the ACSE). Each local parameterization in the atlas is concretely generated by the exponential transformation of a two-body anti-Hermitian operator, providing a map between the Euclidean parameter space and the space of unitary transformations.
Thus, the optimization is no longer defined by a fixed reference set of parameters but rather by a local parameterization at each iteration $n$:
\begin{align}
 \min_{\hat{P}_n \in \mathbb{R}^\mu} \langle \Psi_{n}| e^{-\hat{P}_n} \hat{H}  e^{\hat{P}_n}| \Psi_{n}   \rangle, ~~ ||{P}_n|| < \delta,
\end{align}
where $\mu$ denotes the dimension of the two-body operator space, and the norm on ${P}_n$ indicates that we are staying within a neighborhood around the wave function (replacing the $\epsilon$ in previous formulations). Our optimization is then satisfied if we have a $\Psi_n$ such that for all two-body operators;
\begin{equation}
\begin{split}
    0 &= \frac{\partial}{\partial P^{ij;kl}_nq} \langle \Psi_n| e^{-\hat{P}_n} \hat{H}  e^{\hat{P}_n} | \Psi_n \rangle  \\ &= \langle \Psi_n | [\hat{H},\Gam{i}{j}{k}{l}] | \Psi_n \rangle,
    \end{split}
\end{equation}
which implies that we have fulfilled the ACSE. At each iteration $n$, our current state is mapped to a new state through the exponential mapping.  While it might be thought that the restriction of $\hat{P}$ to a set of two-body operators is too restrictive, as clearly the two-body operator space does not parameterize the unitary group, the current approach iteratively constructs higher order excitations from the reference wave function~\cite{Mazziotti2007, Mazziotti2004, Mazziotti2020} (see Section II.E. of Ref. \cite{Mazziotti2007} for a discussion of the ACSE Ansatz). As we discuss in the next section, because the ACSE is solved iteratively, we can construct quasi-second-order algorithms if we choose each ${\hat P}_{n}$ in the ansatz by considering not only the gradient of the current iteration but also the gradients of previous iterations, which contain information about the curvature.

\subsection{Quasi-Second-Order CQE Algorithms}

\label{sec:qso}

Previous algorithms for solving the ACSE use path-following\cite{Mazziotti2007} or descent\cite{Sand2015} algorithms based on the gradient. While these algorithms are robust to reconstruction errors, gradient-descent algorithms are first-order algorithms with generally linear convergence.  To accelerate convergence, we can consider choosing the search direction $^{2} P_k$ by a second-order approach such as the Newton-Raphson method:
\begin{equation}
^{2} P_k = - {\rm H}_k^{-1}~ {}^2 A_k,
\end{equation}
where $\rm H_{k}$ is the Hessian matrix. Within a certain region of the state space we are guaranteed quadratic convergence\cite{Smith1993}. However, the elements of the Hessian are evaluated according to:
\begin{equation}
\begin{split}
    {\rm H}_n {}^{ij;kl}_{pq;rs} = \frac{\epsilon^2 }{2} \big(\langle \Psi_k | [\Gam{i}{j}{k}{l},[\Gam{p}{q}{r}{s},\hat{H}]] |\Psi_k \rangle \\ + \langle \Psi_k | [\Gam{p}{q}{r}{s},[\Gam{i}{j}{k}{l},\hat{H}]] |\Psi_k \rangle \big),
    \end{split}
\end{equation}
which requires the 4-RDM or its approximation.

To address this issue, we consider the BFGS quasi-Newton method, which uses the Broyden-Fletcher-Goldfarb-Shanno (BFGS) update within Davidon's method, \cite{Davidon1991,Robinson2006} and is summarized in Table \ref{tab:qnQACSE}. At each step of the BFGS method we update an approximate Hessian matrix through a secant equation where the update is designed to keep the Hessian positive definite.  By including a direction based on the approximate Hessian, the BFGS method achieves a superlinear rate of convergence near the solution.

\begin{table}
  \caption{Quasi-Newton CQE algorithm.}
      \label{tab:qnQACSE}
  \begin{ruledtabular}
  \begin{tabular}{l}
  {\bf Quasi-Newton CQE} \\
  \hspace{0.5in} Set $0 \rightarrow n $ \\
  \hspace{0.5in} {Initialize} $| \Psi_{0} \rangle, {}^2 {A}_0$, and ${B}_0$. \\
  \hspace{0.5in} {Continue until $||{}^2 A_n ||< \delta$}:  \\
  \hspace{1.0in} { {\bf Step 1:} Update ${}^2 {P}_n^{} = {B}^{-1}_n ~  {}^2 {A}_n$ } \\
  \hspace{1.0in} { {\bf Step 2:} $\min_{\alpha_n} E_n(\alpha_n \hat{P})$} \\
  \hspace{1.5in} {$\circ$ } $|\Psi_{n+1}\rangle  = e^{\alpha_n \hat{P}_n} |\Psi_n \rangle  $ \\
  \hspace{1.5in} {$\circ$ } ${s}_n = \alpha_n {}^2 {P}_n $ \\
  \hspace{1.0in} { {\bf Step 3:} Evaluate} $^{2} {A}_{n+1}$ \\
  \hspace{1.5in} {$\circ$ } ${y}_n  = {}^2 {A}_{n+1}-{}^2 {A}_{n}$ \\
  \hspace{1.0in} { {\bf Step 4:} }  Calculate $B^{-1}_{n+1}$ \\
  \hspace{1.0in} { {\bf Step 5:} } $n+1 \rightarrow n $ \\
  \end{tabular}
  \end{ruledtabular}

\end{table}

At any particular iteration, given a wave function $|\Psi_n \rangle$, the ACSE residual $\hat{A}_n$, and the inverse of the approximate Hessian $B_n$, we define the step direction as:
\begin{equation}
{}^2 {P}_n^{ij;kl} =  -\sum_{pqrs} (B^{-1}_{n}) {}_{pq;rs}^{ij;kl}\,~{}^2 A^{pq;rs}_n.
\end{equation}
We next minimize the energy by a line search to obtain a direction $\alpha_n$ which satisfies conditions of sufficient descent and curvature (Wolfe conditions). In our local frame, we have the following auxiliary BFGS functions $s_n = \alpha_n \hat{P}_n$ and $y_n = \hat{A}_{n+1} - \hat{A}_n$. Using these, we calculate $B_{n+1}^{-1}$ according to the BFGS formula:
\begin{equation}
    B_{n+1}^{-1} = (I - \frac{s_n y_n^T}{y_n^T s_n}) B_{n}^{-1} (I - \frac{y_n s_n^T}{y_n^T s_n }) + \frac{s_n s_n^T}{y_n^T s_n}.
\end{equation}
We then increase $n$ and continue until the gradient norm $||{}^2 A||$ satisfies a convergence threshold.  Step~4 in Table~1 can be replaced with a suitable update replacement, and in the results we demonstrate the use of a limited-memory BFGS implementation, denoted $l$-BFGS. \cite{Robinson2006}

In practice, because of the redundancy of the certain elements of the 2-RDM, we can store the ${}^2 A$ matrix as a vector in a compact representation, and then store the $B^{-1}$ matrix exactly. For larger systems, it is likely that even this would be prohibitive, and instead the limited-memory approach would be necessary, where we store only $k$ previous steps, which is equivalent to $k$ 2-RDMs. On a classical device, the algorithm would be very similar, although instead of updating the wave function, we would have a 2-RDM update step. Both the 2-RDM and ${}^2 A$ updates would require the classical evaluation of the ACSE and hence, a reconstructed 3-RDM.

On a quantum computer we have significant advantages in that we are not reconstructing the 3-RDM. This means that we generate (up to statistical and noise-related errors) pure 2-RDMs at each step, and do not have to consider the $N$-representability of the 2-RDM, or the step size in the solution of the differential equations. An important note is that if we want to evaluate the ACSE's residual in the classical part of the algorithm rather than in the quantum part by tomography of an auxiliary state, we must be sure that the residual is sufficiently accurate to estimate the curvature.  While an approximated 3-RDM can give good enough information to obtain chemical accuracy in a number of instances, for rigorous convergence the cumulant portion of the 3-RDM should be measured by tomography. Practically, this would entail alternating evaluations of the 3-RDM (for the ACSE) and the 2-RDM (for energy evaluations).

As an alternative to the quasi-Newton approaches, we can instead use the nonlinear conjugate gradient (CG) approaches.\cite{Robinson2006}  The nonlinear CG method does not require the storage of a Hessian or approximate Hessian, and instead involves only a simple update step governing the contribution of the previous search direction $\hat{P}_k$. A description of the generic CQE algorithm with a CG solver is described in Table \ref{tab:cgQACSE}. There are numerous modifications to the conjugate gradient method, which we do not explore here.\cite{Hager2006} Some such modifications include preconditioning schemes, modified update coefficients, as well as additional criteria on resets and step lengths.

\begin{table}
  \caption{Conjugate gradient CQE algorithm.}
    \label{tab:cgQACSE}
  \begin{ruledtabular}
  \begin{tabular}{l}
  {\bf Conjugate Gradient CQE} \\
  \hspace{0.5in} Set $0 \leftarrow n $ \\
  \hspace{0.5in} {Initialize} $| \Psi_{0} \rangle, {}^2 {A}_0$ \\
  \hspace{0.5in} {While $||{}^2 A_n || >\delta$}:  \\
  \hspace{1.0in} { {\bf Step 2:} $\min_{\alpha_n} E_n(\alpha_n \hat{P})$} \\
  \hspace{1.0in} { {\bf Step 3:} Evaluate} $^{2} {A}_{n+1}$ \\
  \hspace{1.0in} { {\bf Step 4:} }Calculate $\beta_{n+1}$ \\
  \hspace{1.0in} { {\bf Step 5:} }${}^2 {P}_{n+1} =  -{}^2 A_{n+1}+\beta_{n+1}{}^2 {P}_n$ \\
  \hspace{1.0in} { {\bf Step 6:} }$n+1 \rightarrow n $ \\
  \end{tabular}
  \end{ruledtabular}
\end{table}

\subsection{Resource-Optimized CQE Search Directions}

\label{sec:sro}

In light of the current representation of the CQE algorithm, one can see that the resources demanded on the quantum computer will depend heavily on the selection and implementation on the search direction, as each term must be implemented individually. However, any modification to the search direction will be detrimental to the rate of convergence (and can potentially also negate the theoretical results). Thus, we would like to find a tradeoff between potentially reducing the number of terms and preserving a descent direction.\cite{Robinson2006}

We thus focus on two approximations which still preserve the essential nature of the ideal CQE approach: $p$-depth and operator sparsification. First, we introduce a more formal way of describing the ACSE ansatz at a given iteration $n$. Let $\{\hat{\mathcal{P}}_l^{(n)}\}$ , be an ordered set of anti-Hermitian two-body operators:
\begin{equation}
\hat{\mathcal{P}}_l^{(n)} = \sum_{ijkl} {\mathcal{P}}^{(n)}_l {}^{ij;kl}\Gam{i}{k}{l}{j} .
\end{equation}
In the exact CQE approach, each iteration adds a new two-body operator to the set.  In general, we can write the ACSE ansatz at the $N$-th iteration as:
\begin{equation}
|\Psi_n\rangle = \prod_{l}^m e^{\hat{\mathcal{P}}_l^{(n)}} |\Psi_0 \rangle .
\end{equation}
where $m$ is the total number of two-body exponential operators we are implementing.  We define the $p$-depth as follows. Given a search direction $\hat{P}_{n}$, we iterate over the elements $P^{ij;kl}_n$ and the set of operators $\{ \hat{\mathcal{P}}_l^{(n)}\}$ from $l=m-p$ to $l=m$. If an element $P^{ij;kl}_n$ was included in a previous operator, we update that two-body operator as:
\begin{equation}
{\mathcal{P}}_{n-p}^{ijkl} \leftarrow  {\mathcal{P}}_{n-p}^{ijkl}+{P}_n^{ijkl}.
\end{equation}
From the definition, the element will only be in one of the $p$-previous operators. If elements cannot be assigned to a previous $\mathcal{P}^{(n)}_l$, a new operator is appended. The ACSE only provides information on the gradient around the exterior of the wavefunction, and because in general each iteration does not commute with previous iterations for $p>0$, this can be considered an approximate scheme of implementing the CQE. As an example, if all terms are included in the initial $\hat{\mathcal{P}}$, then we have a single exponential form in the approximate linear region, and any $p$-depth greater than 1 will be equivalent. Note that such a case is similar to a generalized UCCSD ansatz with different ordering under the first-order trotterization; however, the method of updating does not reflect the true gradient terms, and as such provides an approximation of the operator in the linear region.

The second approach is an operator sparsification scheme, which effectively reduces the number of new terms appended at each step. The method of ordering is of particular importance, and we investigate two options. First, we can sort the elements according to their absolute value, which is important for implementation purposes (i.e., we can remove the smallest elements first). Alternatively, we can sort elements according to the energy contribution in the descent direction. This value is obtained element-wise as the product $A_n^{ij;kl}P_n^{ij;kl}$.  For gradient-descent approaches, these two criteria are equivalent, and in previous work only the former approach was used.

We can also control the number of terms that are removed through a constant $c \in [ 0,1 ] $. This is a constant scaling factor where $c=1$ (strict truncation) indicates only the largest term in $\hat{P}$ is included and $c=0$ (no truncation) indicates that all terms in $\hat{P}$ are included. Another potentially useful control instead of a constant scaling factor (which we do not explore here) would be to limit explicitly the number of terms included in each term. Finally, we have an $\textsc{include}= \{ {\rm True, False} \}$ option whereby elements of the search direction that would be assigned to the previous $p$-operators due to the $p$-depth specified, are included in the sparsification scheme.

\section{Applications and Results}

In this section we look at applications of the optimized CQE scheme. First, we investigate the role of different optimizers. Second, We present discuss schemes related to practical implementations and modifications to the search direction.  Third. we show results with a CQE utilizing the unencoded ACSE, and finally, we compare our results to variational quantum eigensolvers, including the ADAPT-VQE algorithm.

\subsection{Implementation of Optimized CQE}

We begin by comparing the exact (up to a first-order trotterization) implementations of different methods for the ${\rm H}_4$ system at different bond lengths, corresponding with differing degrees of electron correlation. In this work our convergence criteria is typically taken to be the Frobenius norm of the ${}^2A $ matrix. For the conjugate gradient approach, we do not utilize any preconditioning, and use the update strategies of Fletcher and Reeves.\cite{Robinson2006} The limited BFGS strategy utilizes 3 previously stored steps. Figure~\ref{fig:opt_qacse} displays our results. Further computational details are included in the Appendix.

\begin{figure*}
    \centering
    \includegraphics[scale=0.6]{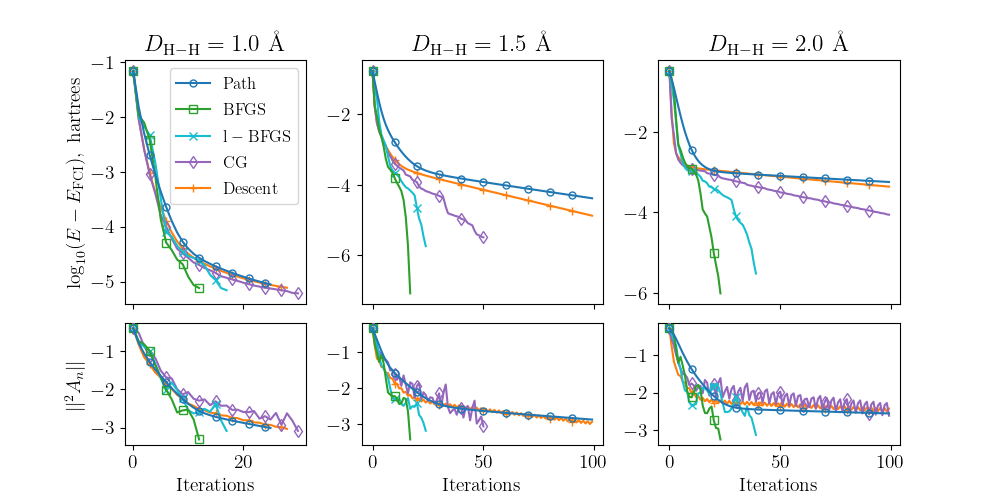}
    \caption{Comparison of methods for generating $\hat{P}$ at different ${\rm H-H}$ distances for ${\rm H}_4$. Near equilibrium, similar patterns can be seen, but away from equilibrium, gradient-descent-based approaches slow down considerably. In particular, when optimizing the 1-d step size according to a quadratic fit, an oscillatory pattern can be seen in the norm of ${}^2 A$, indicating a potential valley in a direction between the oscillating gradients. The conjugate gradient approach allows for slightly more flexibility but exhibits stronger oscillations in the gradient. The quasi-Newton approaches offer quick convergence across all regions of dissociation.}
    \label{fig:opt_qacse}
\end{figure*}

While for the bond distance near equilibrium $D=1$~\AA~, the different optimizers in the ACSE appear to have no apparent advantage, the more correlated distances show strong deviations between the different approaches. In particular, the $1.5$~\AA~case, the convergence flattens when gradient-descent based approaches are used. For $2.0$~\AA~this is accentuated, as we appear to have entered a region where the gradient is quite shallow. The quadratic step appears to be even slightly worse in optimizing the norm of the ${}^2 A$ matrix than a simple gradient descent. The conjugate gradient approach appears to be better but shows strong oscillations in the gradient norm (note if we loosen the update parameter, these oscillations decrease, but we do not observe a significant increase in convergence). The most successful approaches are the quasi-Newton methods, which are able to achieve high accuracy results in only a few iterations. The $l$-BFGS offers a reliable approach as well, with a quality between the conjugate gradient and full BFGS methods.

While this approach is the closest to the ideal implementation, and offers some advantages for a classical quantum simulation (i.e., similar to the classical ACSE, except instead of the 2-RDM we only need to store the statevector), for near-term applications there are several constraints. In a noiseless regime (note, the importance of noise is also relevant, but as this also changes the optimization strategy, is not addressed directly here), addressing the compactness or efficiency is an important problem. In particular, we would however like to know if we can reduce the amount of terms that are added at each iteration, using ideas mentioned above. These result in an appproximate search direction, which we constrain to represent a descent direction.

We use the $1.5$ \AA~case, which contains nontrivial electron correlation and starts to differentiate between different optimizers, to look at ways we can modify the search direction. Table~\ref{tab:approx_line_bfgs} explores the number of iterations and CNOT gate cost for a variety of options with the BFGS optimizer. In particular, we examine the absolute norm or energy contribution for the sparsity operator acting on the search direction (upper and lower quadrants), as well as the inclusion or exclusion of terms that appear in the $p$-depth addition scheme (specified by $\textsc{include}$). For each of these criteria, we look at different $p$-depths and values of the $\textsc{sparse}$ scheme.

\begin{table*}
    \centering
      \caption{Reported number of iterations and CNOT gates (in brackets, $\times 10^3$) for approximate implementation schemes for the CQE with BFGS optimizer with linear H$_4$ at a distance of $1.5$ \AA . We varied the inclusion of terms in the selection of $P$ (sorted by either $|P^{ij;kl}|$, or by the descent condition), the manner of truncation of $P$, as well as varying the sparsity  and the $p-$depth for each condition. $*$ represents a run that did not converge, and $\times$ indicates runs that reached the maximum iterations (300).}
    \label{tab:approx_line_bfgs}
    \begin{tabular}{c|ccccc|ccccc|ccccc|ccccc}
    & \multicolumn{10}{c|}{$|P_n^{ij;kl}|$}  &  \multicolumn{10}{c}{$(A_n P_n)^{ij}_{kl}$}  \\
   & \multicolumn{5}{c}{$\textsc{include}=\textrm{False}$} & \multicolumn{5}{c|}{$\textsc{include}=\textrm{True}$}  &  \multicolumn{5}{c}{$\textsc{include}=\textrm{False}$} & \multicolumn{5}{c}{$\textsc{include}=\textrm{True}$}  \\
   \hline\hline
  \multirow{2}{*}{ $\textsc{sparse}[c]$} &\multicolumn{5}{c|}{$p-$depth} & \multicolumn{5}{c|}{$p-$depth} & \multicolumn{5}{c|}{$p-$depth} & \multicolumn{5}{c}{$p-$depth}
   \\
 & 9 & 7 & 5 & 3 & 1  &  9 & 7 & 5 & 3 & 1  &  9 & 7 & 5 & 3 & 1  &  9 & 7 & 5 & 3 & 1   \\
            \hline
%
 \multirow{2}{*}{0.9} &
 \multirow{2}{*}{*} & \multirow{2}{*}{*} & \multirow{2}{*}{*} & \multirow{2}{*}{*} & \multirow{2}{*}{*} &
 76 & 73 & \multirow{2}{*}{*} & 65 & 117 &
 \multirow{2}{*}{$\times$} & \multirow{2}{*}{$\times$} & 283 & \multirow{2}{*}{$\times$} & \multirow{2}{*}{$\times$} &
 47 & 48 & 60 & 80 & 107
 \\
 &
 & & & & &
 [1.8] &  [2.5] & & [4.0] & [8.9] &
& & [18] & & &
[1.6] & [2.2] & [4.0] & [4.0] & [8.1]
 \\
 %
  \multirow{2}{*}{0.5}  &
 \multirow{2}{*}{*} & \multirow{2}{*}{*} & \multirow{2}{*}{*} & \multirow{2}{*}{*} & \multirow{2}{*}{*} &
 25 & 25 & 42 & \multirow{2}{*}{*} &  \multirow{2}{*}{*} &
 71 & 139 & 123 & 119 & 75 &
 26 & 35 & 35 & 39 & 63
 \\
  &
 & & & & &
 [0.90] & [0.90] & [2.6] & & &
 [0.90] & [5.4] & [6.6] & [7.6] & [13] &
 [0.91] & [1.9] & [2.5] & [3.3] & [7.6]
 \\
 %
 \multirow{2}{*}{0.25}  &
 26  & 28  & 28  & 27  & 31 &
 24 & 24 & 24 & 28 & 29 &
 63 & 67 & 70 & 81 & 76 &
 26 & 26 & 30 & 36 & 46
 \\
 &
 [1.6] & [3.4] & [4.6] & [5.8] & [11] &
 [1.3] & [1.3] & [2.2] & [6.5] & [11] &
 [0.93] & [0.93] & [3.7] & [6.1] & [15] &
 [1.2] & [1.2] & [3.3] & [5.1] & [8.8]
 \\
 %
  \multirow{2}{*}{0.125}  &
 22 & 22 & 22 & 23 & 24 &
 21 & 21 & 22 & 21 & 24 &
 47 & 56 & 74 & 44 & 54 &
 24 & 26 & 26 & 29 & 38
 \\
 &
 [1.5] & [1.5] & [1.5] & [7.0] & [13] &
 [1.5] & [1.5] & [5.6] & [6.5] & [13] &
 [1.4] & [3.9] & [6.3] & [7.8] & [16] &
 [1.5] & [3.2] & [3.7] & [5.5] & [8.8]

    \end{tabular}

\end{table*}

\begin{figure}[t]
    \centering
    \includegraphics[scale=0.6]{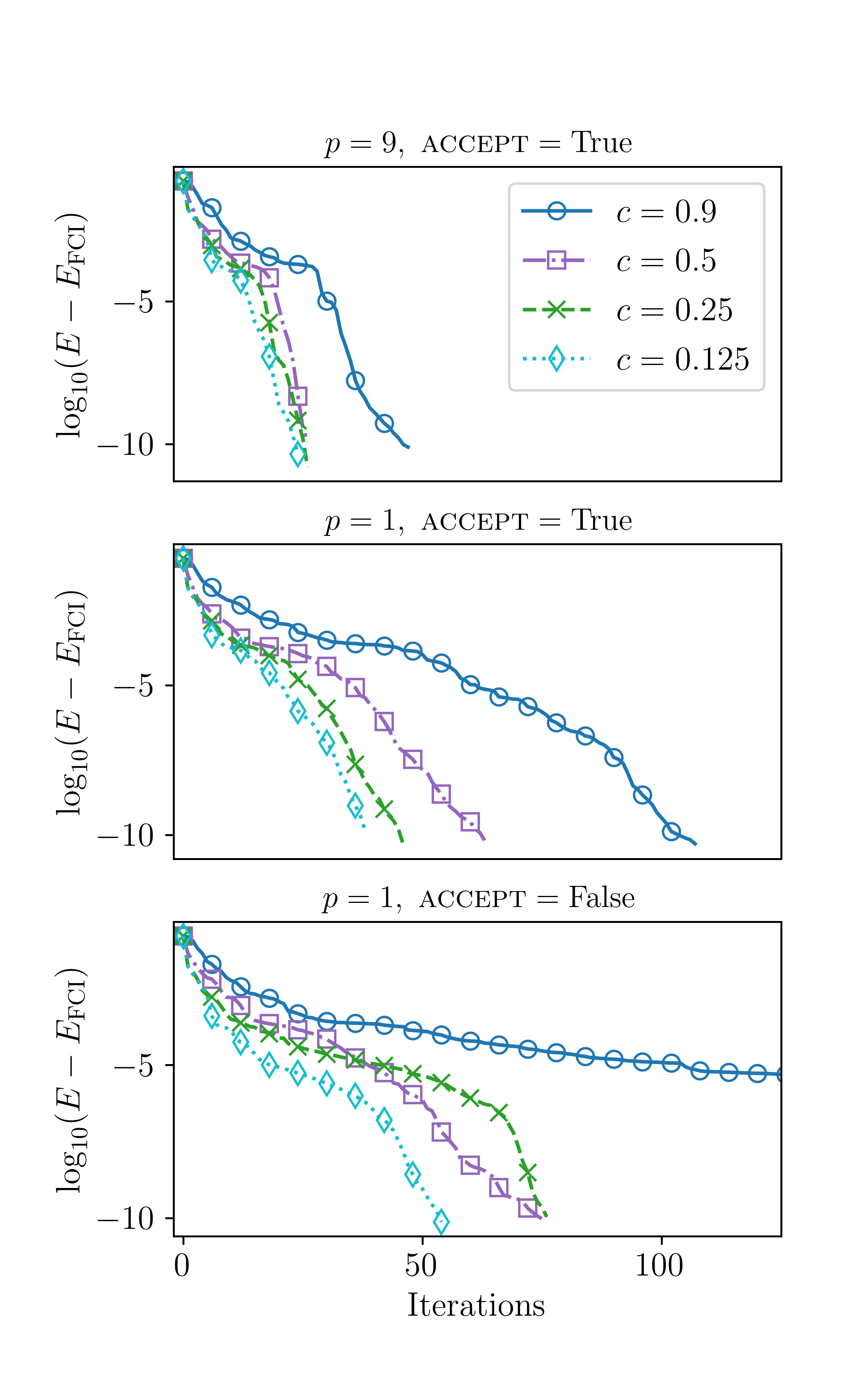}
    \caption{Low threshold ($||{}^2 A||=10^{-5}$) results for varying values of sparsity with different $p-$depths and acceptance criteria, utilizing the BFGS optimizer for molecular ${\rm H}_4$ at the intermediate distance ($D=1.5$~\AA), with the $\textsc{descent}$ condition. In the top two plots, we use $\textsc{accept}=\textrm{True}$, where elements below the sparsity threshold are automatically included if they are in included in one of the previous $p-$terms. The bottom plot has $\textsc{accept}=\textrm{False}$. Only the first 125 iterations are shown.}
    \label{fig:approx_line_bfgs}
\end{figure}

A number of interesting trends emerge. First, there is a difference in application of the sparsification operator acting on elements according to their energy contribution or absolute value. Namely, when using large $c$ for the absolute value, problems in the optimization can occur. Namely, these are instances where the search direction has little overlap with the gradient, and the largest term selected is ordered in such a way that it is not strictly increasing. The descent condition however, is able to converge across every configuration, albeit at different rates of convergence. This also leads to a stratification in the rates of convergence for the $\textsc{descent}$ condition, which can be seen in Figure~\ref{fig:approx_line_bfgs}. While the absolute value condition leads to accelerated convergence in almost all instances, it does seem more sensitive to using restricted operators. Of course, the inclusion of previous terms does seem to alleviate this problem, and because it seems to occur when the search direction is nearly orthogonal to the gradient, resetting the optimization might allow the optimization to continue. It is also possible that the order of magnitude for the descent condition should be lower than the absolute value condition, although this could vary significantly based on the system.

Second, the $\textsc{include}$ option has a strong impact on the rate of convergence. For both selection criteria, inclusion of previous terms clearly helps in assisting the overall convergence. As this can be considered as a way of increasing the pool of operators at each step, depending on both $c$ and the $p$-depth, the advantage here makes sense. Third, the $p$-depth appears to have a two-fold role. First, and more generally, it serves to reduce the total number of terms needed in the ansatz. That is, as the $p$-depth increases, the number of total terms in the ansatz decreases. Additionally, for a given $c$, we do see numerous instances where the total number of iterations decreases as the $p$-depth increases when $\textsc{include}={\rm True}$. When $\textsc{include}={\rm False}$, the trends are a bit unclear, and the optimization tends to be more sensitive. Interestingly, for $c=0.5, 0.25, $and $0.125$, with $\textsc{include}={\rm False}$ the total iterations appear to increase with increasing $p$-depth, and then decrease. More sparse (see $c=$0.9, 0.5) truncations result in slower convergence, which can be aided with the $\textsc{include}$ option, but not completely mitigated.

While these results are obviously not generalizable to every system, it is likely that some of these trends can be seen elsewhere. We expect that the $p$-depth can lead to lower circuit depths. Additionally, while sparser operators are desirable from a NISQ perspective, optimization with respect to a single parameter is clearly detrimental to the rate of convergence. This can be mitigated through expanding the pool directly with more terms and the sparsification operator, or indirectly through the $p$-depth.

\subsection{Encoding-Free CQE}

The encoding-free (or unencoded CQE) approaches for preparing states as an alternative to fermionic state preparation have recently been explored\cite{Ryabinkin2018, Xia2020, Tang2019, Yordanov2020},  within the VQE framework as well as in attempting to understand the success of heuristic and non-fermionic ansatz preparation. Recent works by the present authors showed that the fermionic 2-RDM can be functionalized from a qubit-particle wavefunction.\cite{Mazziotti2021} As long as fermionic tomography is performed on a $N$-qubit particle state, the 2-RDM represents a valid fermionic 2-RDM. Additionally, in recent work we introduce an encoding-free CQE algorithm which evaluates the anti-Hermitian component of the two-qubit-particle contraction onto the Schr{\"o}dinger equation.
Table~\ref{tab:cqe_cost} displays calculations for a number of bond distances of ${\rm H}_5$ for both the encoded and unencoded CQE  cases with the BFGS algorithm as an example system.

\begin{table}[]
    \centering
        \caption{Comparisons of total iterations, total CNOT gate count, and average CNOT gates per iteration for unencoded and encoded CQE using the BFGS optimizer for ${\rm H}_5$ at various bond distances from equilibrium in the minimal STO basis. The accuracy of both approaches is largely similar across the dissociation curve, though for $+1.00$ and $+1.25$ distance separation, the unencoded CQE requires more iterations.}
    \label{tab:cqe_cost}
    \begin{tabular}{c|ccc}
    \multirow{1}{*}{$D - D_{\rm eq}$} & \multirow{1}{*}{Iterations} &  Total CNOT $\times 10^{4}$ & $\langle \rm CNOT_k \rangle  \times 10^2$ \\
\AA  & (CQE,UCQE)  &   (CQE,UCQE) & (CQE,UCQE) \\ \hline \hline
    $-0.25$ & 26, 33 & 1.7, 1.5  & 6.4, 4.5 \\
    $+0.00$ & 29, 36 & 2.1, 1.8  & 7.4, 5.0\\
    $+0.25$ & 47, 49 & 3.8, 2.7  & 8.2, 5.5 \\
    $+0.50$ & 40, 34 & 4.3, 2.6  & 11., 7.8\\
    $+0.75$ & 39, 28 & 5.7, 2.8  & 15., 9.9 \\
    $+1.00$ & 42, 43 & 6.9, 4.4  & 16., 10. \\
    $+1.25$ & 47, 80 & 8.0, 7.4  & 17., 9.3 \\
    \end{tabular}

\end{table}

The unencoded CQE under optimization matches the fermionic case in most instances, and consistently has a smaller average number of CNOT gates per iteration. For the two longest separation lengths, the number of iterations required does increase, leading to similar CNOT counts for the total ansatz. A future goal would be to incorporate compilation schemes or adjust the set of ACSE or unencoded ACSE excitations to favor a largely commuting pool.

\subsection{Comparison with VQE}

\begin{figure}
    \centering
    \includegraphics[scale=0.55]{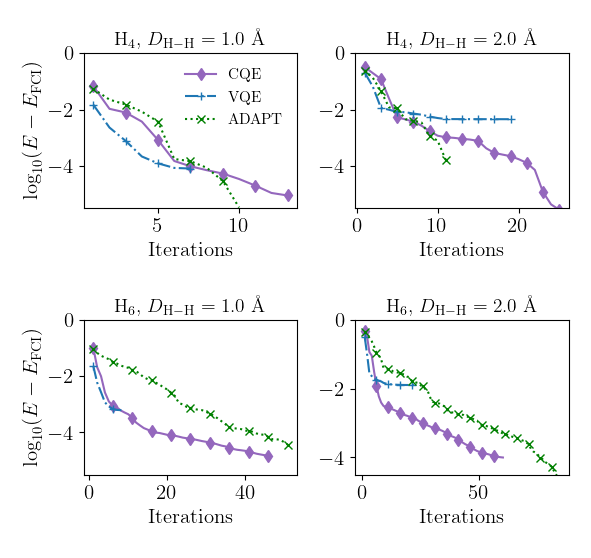}
    \caption{Simulations of molecular $\rm H_4$ and $\rm H_6$ for separations of 1~\AA,  and 2~\AA~ utilizing the CQE, VQE, and ADAPT-VQE with a BFGS optimizer. For ${\rm H}_4$ a minimal ansatz is reached in the ADAPT through the VQE subroutine, and for ${\rm H}_6$ the CQE algorithm slightly outperforms the ADAPT. VQE, here using the UCCSD anastz, provides rapid convergence for near equilibrium states, but has significant errors at larger separations, which are overcome by the iterative ADAPT-VQE and CQE algorithms.
    }
    \label{fig:h4h6}
\end{figure}

Finally, we compare the CQE approach utilizing a BFGS optimization with other known quantum algorithms. While in previous work\cite{Smart2021_benzyne} similarities between iterative nature of the ACSE and ADAPT-VQE were discussed, here, we provide example calculations of VQE, ADAPT-VQE, and the ACSE that demonstrate fundamental differences in these algorithms. These are included in Figure~\ref{fig:h4h6}, as well as in Table ~\ref{tab:gradient_evaluations}.

While the VQE results in Figure~\ref{fig:h4h6} are not that surprising based on the use of the unitary coupled cluster ansatz, we still can see some interesting comparisons. For equilibrium distances, UCCSD provides a good ansatz, and there are numerous methods exploring the UCC ansatz \cite{Romero2017,Lee2019}. By comparison, with ADAPT we are able to obtain seemingly arbitrary convergence, matching previous work. However, it is worth noting that the iterative cost of the ADAPT is much higher than either VQE or the CQE. The CQE on the other hand performs quite well in a variety of instances, with the most challenging case being dissociated ${\rm H}_6$, where higher order excitations dominate and the system is strongly correlated. While the number of macro iterations for the ADAPT procedure might look only slightly worse than the CQE or VQE approaches, when taken into account with the VQE cost, (i.e. micro + macro iterations), the length of the ADAPT procedure is somewhat unwieldy, namely due to the VQE subroutine. The total number of gradient and residual evaluations for each of these instances is seen in Table \ref{tab:gradient_evaluations}.

\begin{table}[]
    \centering
        \caption{Comparison of gradient element (VQE) and residual element (ACSE) evaluations for the CQE, VQE, and ADAPT-VQE methods, corresponding to simulations for molecular $\rm H_4$ and $\rm H_6$ at separations of 1~\AA,  and 2~\AA. Note the ADAPT-VQE here has a symmetry adapted pool of operators, which are not implemented here in the CQE or VQE approaches. The VQE tolerance is also taken to be quite low, i.e. $10^{-3}$ in the norm of the parameter vector (whereas the VQE subroutine in the ADAPT procedure is generally lower).}
    \begin{tabular}{cc|cccc}
     Method & Quantity  & ${\rm H}_4$, 1~\AA & ${\rm H}_4$, 2~\AA & ${\rm H}_6$, 1~\AA & ${\rm H}_6$, 2~\AA\\ \hline \hline
          \multirow{2}{*}{CQE} & Iterations & 13 & 25 & 46 & 60 \\
      & Residuals & 1950 & 3750 & 38640 & 50400 \\
      \hline     \multirow{2}{*}{VQE} & Iterations &7 & 19 & 9 & 22 \\
    &  Gradients & 182  & 494 & 1053 & 2574 \\ \hline
   \multirow{3}{*}{ADAPT-VQE}   &  Macro & $10$ & $12$ & $51$ & $84$ \\

      &  Micro & 71 & 167 & 1221 & 5854 \\
      & Gradients & 362 & 1226 & 43112 & 354988 \\
      & Residuals & 660 & 792 & 16830 & 27720
    \end{tabular}
    \label{tab:gradient_evaluations}
\end{table}

We can also look at qubit implementation of the ADAPT scheme, which here follows the qubit-particle excitation based scheme of Yordanov et al. \cite{Yordanov2020}, and not the quasi-particle approach taken by the original authors. We find a similar result to previous work, namely that for stretched LiH conserving the particle number and projected spin leads to essentially the same qubit-based excitations. The main difference seen (as a result of the relative scale mostly) in these results is for the CNOT cost of the unencoded CQE approach, although a similar decrease in the CNOT cost of the IQEB approach exists as well. While the approaches appear to be inversely related in the rate of convergence (through total iterations), and the CNOT count (where the CQE schemes are more costly), the number of parameter evaluations differs substantially. While here we do not distinguish between the residuals of the ACSE as parameters and the VQE parameters, as these can be obtained a number of different ways (for instance, the ACSE residuals can be taken from either the 2-RDM with a quantum solver or the 3-RDM on the quantum computer, and numerous methods of measuring VQE gradients exist as well), for larger system limits on the number of parameters should be considered.

\begin{figure}
    \includegraphics[scale=0.55]{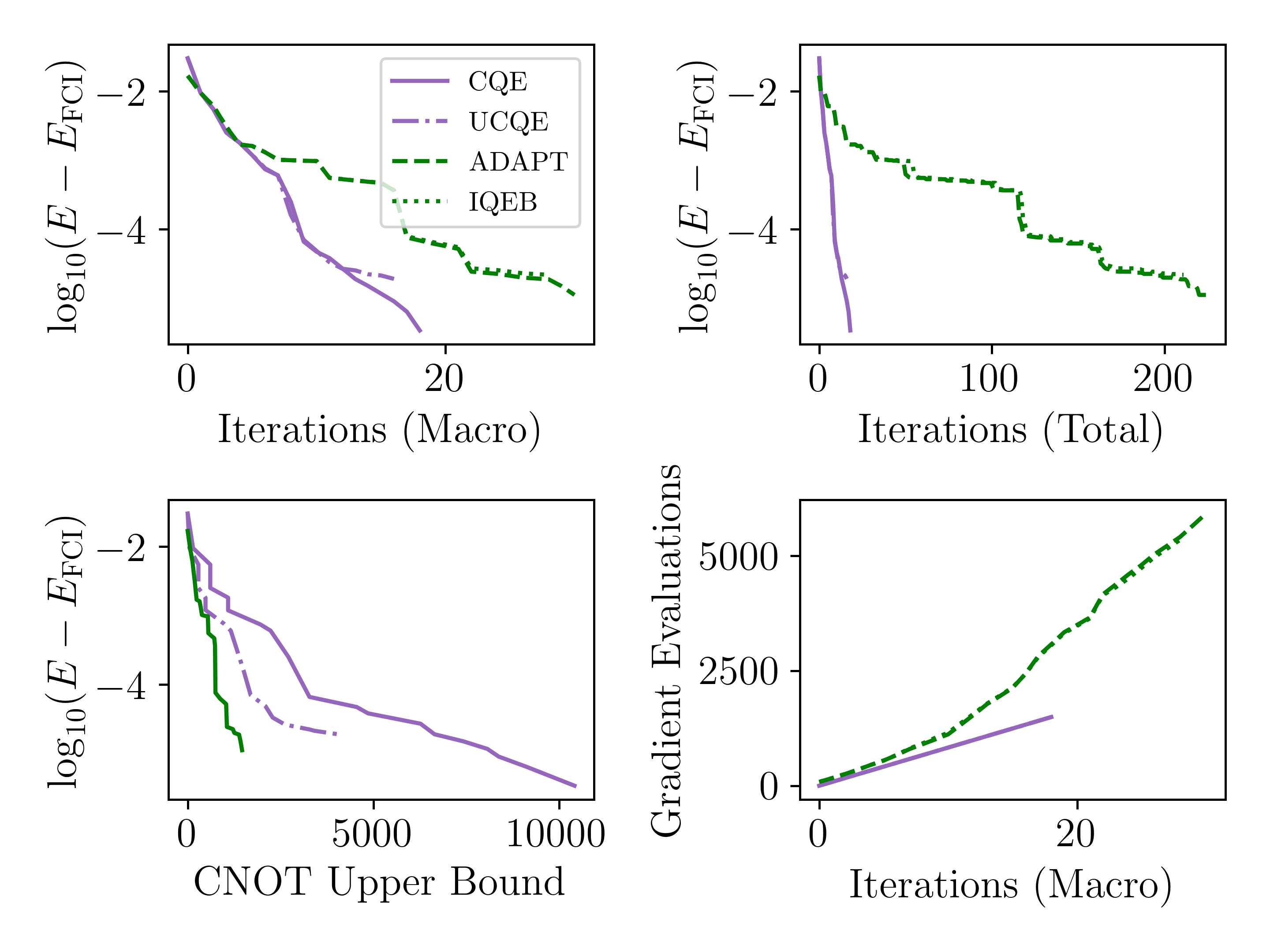}
    \caption{Simulations of molecular lithium hydride at a separation of 2~\AA~utilizing CQE, VQE, fermionic ADAPT-VQE and the iterative qubit-excitation based ADAPT-VQE (IQEB), essentially a unencoded APAPT-VQE algorithm. The upper-left  shows the macro iterations of each scheme, which are nearly identical for the encoded and unencoded forms, and are only slightly slower for the ADAPT. The upper-right shows the total iterations, including micro iterations in the VQE procedure with a threshold of $10^{-3}$ in the parameter vector. The lower-left compares the energy convergence with the ansatz cost, showing that the ADAPT does produce a more compact representation of the ansatz. The lower-right shows the number of parameter evaluations (a lower bound on the number of gradients evaluated at each optimizer step), which is linear scaling with the ACSE procedure but for ADAPT becomes quadratic with respect to the number of iterations.}
    \label{fig:lih}
\end{figure}

Despite achieving quicker convergence, from all of these examples we see that the main drawback to the CQE algorithm is the iteratively increasing CNOT cost, which is due to the use of additional gates at each step. While this should be reasonable for high performing quantum devices, for near-term devices further reduction of the CNOT count is important.  However, the gain in performance we see by performing a quasi-second order optimization in the local parameter space is quite significant. Additionally, when compared to CNOT gates of the gradient-descent-based methods (as in Fig.~\ref{fig:opt_qacse}), the optimized-ACSE allows for more flexibility in constructing compact ansatz.

\section{Discussion}

While generic algorithms have been known for approaching the problem of quantum simulation for a while, calculations involving increasingly complex systems have only recently begun to emerge. These require the advancement and development of new quantum algorithms, similar to the past century of classical quantum chemistry algorithms. The CQE offers an approach which is potentially beneficial in the near-term, and could potentially solve some of the shortcomings of VQE.

From the discussion in section \ref{sec:para}, we can see that the variational principle used in the VQE and in the current CQE algorithm are similar in that they solve an optimization problem, but differ in the goal of the minimization. In the VQE we are often trying to minimize the energy of a state through a global parameterization, and thus finding a suitable unitary transformation is one issue. Barren plateaus or regions where the optimization fail become likely with an increasingly large parameter space. Additionally, the suitability (i.e. over or under paramterization) of our state is often in question. In the ACSE algorithm, the local parameterization which is more suitable for RDM representations leads to a locally updated optimization. Importantly, in a VQE the optimization converges \emph{towards a solution of the VQE problem}, which is not the ACSE. As the VQE subproblem becomes larger and larger (i.e. in adaptive schemes), eventually the VQE solution can (but by design will not necessarily) satisfy the ACSE. In the CQE approach presented here, we have convergence \emph{towards our contracted eigenvalue problem}, and not a variational subproblem.

From our calculations we can also understand some elements of the ADAPT-VQE algorithm as is related to the CQE algorithm. The ADAPT-VQE method chooses the largest ACSE residual at each macro iteration. This leads to a flexible and efficient ansatz, that when not restarted, will by construction improve the energy in the VQE. However, the re-optimized state is often not close to the previous state, highlighting the strong variational nature of the ADAPT algorithm. Restarting the VQE optimization, which has been done in some ADAPT work, can lead to a suboptimal solution of the ACSE, or for the VQE subroutine to fail. Recent work by Liu et al.\cite{Liu2020} used a reconstructed 3-RDM in the ACSE to obtain approximate residuals to seed the ADAPT algorithm. Because these approximate gradients can differ in a substantial way from the exact gradients, more terms are needed, which significantly increases the variational flexibility of the ansatz. As a result, these calculations exhibited faster convergence and required fewer iterations than traditional ADAPT-VQE.

In this work, the effective number of parameters is kept to the size of the 2-RDM. However, unlike Liu et. al,\cite{Liu2020}, it is important \emph{not} to use the classical ACSE with reconstructed residuals from a reconstructed 3-RDM, as these will likely lead to convergence issues. However, the classical 3-RDM can still be obtained relatively easily on a quantum computer through a variety of techniques. Work involving qubit-particle approaches also shows promise, with another advantage being the increased number of commuting terms which exists between qubit-particle excitations as opposed to fermionic-particle excitations.

Another element which can be overlooked is that the ACSE is not necessarily equivalent to the CSE except when higher order excitations are incldued. Despite this limitation, in the exponential form of the ACSE, higher order excitations can be seen to emerge naturally through products of exponential two-body operations. Additionally, by considering information on the curvature of the space beyond the gradient, we also should include contributions from triple and higher excitations in our selection of operators to propagate the wavefunction. In practice, use of the CQE for solving the ACSE leads to a highly accurate solution.

Despite these benefits, there still are drawbacks with the CQE algorithm. The primary drawback when compared to an algorithm such as the ADAPT-VQE is the large number of CNOT gates. Even with low error CNOT gates, efficient and noise-robust means of obtaining accurate gradients and 2-RDMs will be necessary. All of these also affect the success or failure of the underlying optimization algorithm, and so exploring noise-tolerant approaches will also be critical for near-term applications.

\section{Conclusion}

In this work we address the convergence of the contracted quantum eigensolver using tools from optimization theory. By using methods beyond traditional gradient descent, we achieve superlinear convergence, allowing us to propagate the wavefunction rapidly towards a solution of the ACSE. Practical implementations where the search direction is modified to conserve quantum resources show promising reductions in the cost of the algorithm, and we expect further simplification schemes to be attempted aimed at improving the efficiency of the CQE approach. Additionally, the present work provides a basis for understanding approaches which use the ACSE in pool selection, and could lead to further hybrid optimization schemes for use in NISQ applications.

\begin{acknowledgments}
D.A.M. gratefully acknowledges the Department of Energy, Office of Basic Energy Sciences, Grant DE-SC0019215 and the U.S. National Science Foundation Grant No. CHE-2035876.
\end{acknowledgments}

\appendix

\section{Computational Details}
All calculations were performed using the \textsc{hqca} (v22.4)\cite{Smart_hqca_-_hybrid} set of tools, which utilizes \textsc{qiskit} (v0.29.0)\cite{Qiskit} and \textsc{pyscf} (v1.7.6)\cite{Sun2017} for interfacing with quantum simulators and obtaining electron integrals for circuit based simulations. Each simulation utilizes a minimal-basis Slater-type orbital representation (i.e. STO-3G). The Jordan-Wigner (or qubit-particle) mapping was utilized, with parity $\mathbb{Z}_2$ symmetries removed for the majority of examples. Statevector or unitary simulations with no noise were used for each run. The ADAPT-VQE results in Fig~\ref{fig:h4h6} were obtained using code from the respective publication\cite{Grimsley2018}, where analytical recursive solutions of the VQE gradients are used. The threshold for the VQE subroutine in those instances was $10^{-6}$. In the lithium hydride case, code from \textsc{hqca} was used. Parameters for the VQE optimization in the ADAPT-VQE optimizations were not reset in between runs - a single parameter is essentially appended to the parameter vector.

The line-search implementation used in the BFGS, nonlinear CG, and $l-$BFGS optimizations follows from the Nocedal algorithms\cite{Robinson2006}, which is present in the $\textrm{scipy}$ implementation. For some steps (notably the first few steps), often $\alpha_0=1$ is too large, and $\alpha_0 =0.5$ is preferred. While a dynamic step-size is not necessary for BFGS, after the first step we interpolate the $\alpha_0$ with a quadratic based on the current energy, previous energy, and previous gradient information, and then constrain $\alpha_0 \in [0.5,1]$, which we expect in some instances leads to a more appropriate step size.

\bibliographystyle{achemso}
\bibliography{
    biblio/0_general_qc.bib,
    biblio/1_opt.bib,
    biblio/2_acse.bib,
    biblio/3_rdm.bib,
    biblio/4_software.bib,
    biblio/5_adapt.bib,
    biblio/6_cse.bib,
    biblio/7_books.bib,
    biblio/8_vqe.bib
    }
\baselineskip24pt

\maketitle
\end{document}